\documentclass[aps,prl,twocolumn,groupedaddress,showpacs]{revtex4}
\usepackage{epsf}

\newcommand{\vk}{{\bf k}}
\newcommand{\phase}{\rho}
\newcommand{\current}{J}
\newcommand{\intensity}{I}

\begin{document}

\title{Wavefunction statistics in open chaotic billiards}

\author{Piet W. Brouwer}
\affiliation{Laboratory of Atomic and Solid State Physics, 
Cornell University, Ithaca, New York 14853-2501}
\date{\today} 

\begin{abstract}
We study the statistical properties of wavefunctions in a chaotic
billiard that is opened up to the outside world. Upon increasing the
openings, the billiard wavefunctions cross over from real to complex.
Each wavefunction is characterized by a phase rigidity, which is
itself a fluctuating quantity. We calculate the probability
distribution of the phase rigidity and discuss how phase rigidity 
fluctuations cause
long-range correlations of intensity and current density. We also find
that phase rigidities for wavefunctions with different
incoming wave boundary conditions are statistically correlated.
\end{abstract}


\maketitle

Microwave cavities have been used as a quantitative experimental
testing ground
for theories of quantum chaos \cite{kn:stoeckmannbook}. In quasi 
two-dimensional cavities, the
component of the electric field perpendicular to the surface of the
cavity satisfies a scalar Helmholtz equation that is formally
equivalent to the Schr\"odinger equation. Complex field patterns,
which model the wavefunction of an electron in a magnetic field, can
be obtained making judicious use of magneto-optical effects
\cite{kn:so1995,kn:stoffregen1995}. Alternatively, complex
``wavefunctions'' can be observed as travelling waves in open microwave
cavities \cite{kn:pnini1996,kn:barth2002}. Measured distributions of
real and complex wavefunctions in microwave cavities with chaotic ray
dynamics, where, traditionally, ``complex''
means that time-reversal symmetry is fully broken and the phase of the
wavefunction has no long-range correlations, agree with a theoretical 
description in terms of a random superposition of plane waves 
\cite{kn:berry1977}, as well as with random matrix theory
\cite{kn:guhrreview} and supersymmetric field theories
\cite{kn:efetovbook}.

Recently, it has become possible to study the full real-to-complex
crossover using microwave techniques \cite{kn:barth2002,kn:chung2000}. 
The crossover regime is 
qualitatively different from the ``pure'' cases of real or fully
complex wavefunctions. Unlike in the pure cases, the statistical 
distribution of wavefunctions in the crossover regime depends
on the choice of the ensemble: whether variations are taken with
respect to the coordinate ${\bf r}$, the frequency $\omega$, or
both. 
Whereas the theoretical work has been roughly equally divided
between the two approaches, experiments usually need
the additional average over frequency to obtain sufficient statistics
\cite{kn:so1995,kn:stoffregen1995,kn:wu1998,kn:chung2000} (see,
however, Ref.\ \onlinecite{kn:barth2002} for an exception). 

In general, a complex wavefunction may be written as
\begin{equation}
  \psi({\bf r}) = e^{i \phi} (\psi_{\rm r}({\bf r}) + i  
  \psi_{\rm i}({\bf r})), \label{eq:wave}
\end{equation}
where $\psi_{\rm r}$ and $\psi_{\rm i}$ are orthogonal but need not have
the same normalization \cite{kn:french1988}.
 The ratio of $\psi_{\rm r}$ and $\psi_{\rm i}$ is parameterized in terms of 
the normalized scalar product of $\psi$ and its time-reversed, 
\begin{eqnarray}
  \phase &=& 
  \frac{\int d{\bf r}\,  \psi({\bf r})^2}{
  \int d{\bf r}\, |\psi({\bf r})|^2}
  = e^{2 i \phi}
  \frac{\int d{\bf r}\, |\psi_{\rm r}({\bf r})|^2 - |\psi_{\rm i}({\bf r})|^2}
  {\int d{\bf r}\, |\psi_{\rm r}({\bf r})|^2 + |\psi_{\rm i}({\bf r})|^2}.
  \label{eq:phase}
\end{eqnarray}
The square modulus $|\phase|^2$ is known as the ``phase rigidity'' of
the wavefunction $\psi$ \cite{kn:vanlangen1997}. 
Real wavefunctions have $\phase=1$,
whereas $\phase=0$ if $\psi$ is fully complex, {\em i.e.}, $\psi_{\rm r}$ and
$\psi_{\rm i}$ have the same magnitude.
If the average is taken over the coordinate ${\bf r}$ only, whereas the
frequency $\omega$ of the wavefunction is kept fixed, the wavefunction
distribution follows by describing $\psi_{\rm
  r}$ and $\psi_{\rm i}$ as random superpositions of
standing waves 
\cite{kn:zyczkowskilenz1991,kn:kanzieperfreilikher1996,kn:pnini1996}. 
The resulting wavefunction distribution 
depends parametrically on the phase rigidity $|\phase|^2$.
Using a microwave billiard
with a movable antenna, Barth and St\"ockmann have measured such
a ``single-wavefunction distribution'' and found good
agreement with the theory,
obtaining $\phase$ from an independent measurement \cite{kn:barth2002}.
It is the fact that $\phase$ is different for each wavefunction that
leads to the different results for averages over ${\bf r}$
only and over both ${\bf r}$ and $\omega$.
A calculation of
averages with respect to frequency requires a theory of the 
probability distribution of $\phase$.
Such a full wavefunction distribution, which needs theoretical
input beyond the ansatz that each wavefunction $\psi$ 
is a random superposition of plane waves, was first calculated by Sommers and
Iida for the Pandey-Mehta Hamiltonian from random-matrix theory
\cite{kn:sommersiida1994} and by Fal'ko and Efetov
\cite{kn:falkoefetov1994,kn:falkoefetov1996} for a disordered quantum dot
in a uniform magnetic field.

Fluctuations of the phase rigidity $|\rho|^2$ have been identified 
as the root cause for
several striking phenomena in the crossover regime, such as long-range
intensity correlations \cite{kn:falkoefetov1996} and a non-Gaussian
distribution of level velocities \cite{kn:vanlangen1997}. Further, the
existence of correlations between phase rigidities of different
wavefunctions causes long-range correlations between wavefunctions at
different frequencies \cite{kn:absw}. The experimental
verification of these effects addresses aspects of random
wavefunctions
that have not previously been tested.
The relative magnitude of the phase rigidity fluctuations
is numerically small, leading to long-range wavefunction correlations 
of order of 10 percent or less \cite{kn:falkoefetov1996,kn:absw}.
This could explain why intensity distributions measured by Chung
{\em et al.} could not distinguish between theories with
and without phase rigidity fluctuations \cite{kn:chung2000}.

In this letter, we consider wavefunctions in a
billiard that is opened up to the outside world and calculate the
probability distribution of phase rigidities for this case. Although 
time-reversal symmetry is not broken on the level of the wave equation
itself, it is broken by the fact that one looks at a scattering state
with incoming flux in one waveguide only \cite{kn:pnini1996}. 
As we show here, random
wavefunctions in open cavities also have a fluctuating phase rigidity,
and, hence, exhibit the same variety of phenomena as those in cavities
with broken time-reversal symmetry, while they are much easier to
generate in microwave experiments \cite{kn:barth2002}.
An additional advantage of the
open-billiard geometry is the absence of fit parameters: 
The only parameter entering the
wave-function distribution is the total number $N$ of propagating modes in
the waveguides between the billiard and the outside world, which can
be measured independently. 

Following previous works on this subject, we consider the parameter
regime in which the frequency average is taken over a window 
$\Delta \omega \ll c/L \ll \omega$, where $c$ is the velocity of
wave propagation and $L$ the size of the billiard, and in which the
openings occupy only a small fraction of the billiard's boundary.
It is only in this regime that wavefunctions
have a universal distribution and a description in terms
of a random superposition of plane waves is appropriate. We limit ourselves
to (quasi) two-dimensional billiards, in which the electric field
perpendicular to the billiard plane is identified with the
wavefunction $\psi$ and the Poynting vector with the current density
${\bf j} \propto \mbox{Im}\, \psi^* {\bf \nabla} \psi$ \cite{kn:seba1999}.
A calculation of 
wavefunctions inside an open billiard is complementary to a
transport study, for which one is primarily intersted in the 
relation between amplitudes of ingoing and outgoing waves in the
waveguides attached to the billiard, not in the wavefunction inside
the cavity.
Single-wavefunction statistics in open billiards in the universal regime 
was first considered 
by Pnini and Shapiro \cite{kn:pnini1996} and subsequently by
Ishio {\em et al.} \cite{kn:ishio2001,kn:saichev2002}.
Experimentally, wavefunctions in open billiards were investigated
by Barth and St\"ockmann \cite{kn:barth2002}. 

The key to the calculation of $P(\rho)$ in an open cavity is a 
relation between the scalar products of the in-cavity parts of
scattering states $\psi_{\mu}$ and $\psi_{\nu}$ and the
Wigner-Smith time-delay matrix $Q$ \cite{kn:smith1960}, 
\begin{equation}
  \int_{\rm cavity} d{\bf r} \psi_{\mu}({\bf r}) \psi_{\nu}^*({\bf r})
  = Q_{\mu\nu},\ \
  \label{eq:smith}
\end{equation}
where the scattering states have been normalized to unit incoming
flux. Here the index $\mu=1,\ldots,N$ labels the waveguide and
the transverse mode
from which the field is injected into the cavity.
The time-delay matrix $Q = -i S^{\dagger} {\partial S}/{\partial
\omega}$ is the derivative of the scattering matrix $S$. In order to
calculate the scalar product $\rho_{\mu\mu}$ of the scattering state 
$\psi_{\mu}$ and its time-reversed $\psi_{\mu}^*$, we
perform a unitary transformation $U$ that diagonalizes the
Wigner-Smith time-delay matrix $Q$ and rotates the scattering
matrix $S$ to the unit matrix \cite{kn:bfb1997},
\begin{equation}
  S = U^{\rm T} U,\ \
  Q = U^{\dagger} \mbox{diag}\,(\tau_1,\ldots,\tau_N) U.
\end{equation}
The positive numbers  $\tau_i$, $i=1,\ldots,N$, are the 
``proper delay times'', the
eigenvalues of the Wigner-Smith time-delay matrix. Note that the
incoming modes are transformed according to the unitary
transformation $U$, while the outgoing modes transform according
to $U^*$, as required by time-reversal symmetry. In the
transformed basis, all scattering states are standing waves
and, hence, have $\rho_{jj}=1$. Transforming back to the original basis, we 
find
\begin{equation}
  \phase_{\mu\mu} =  
   \frac{ \sum_{j=1}^{N} U_{j \mu}^2 \tau_j }{\sum_{j=1}^{N} |U_{j \mu}|^2
  \tau_{j}}.
  \label{eq:rhomu}
\end{equation}

\begin{figure}[t]
\epsfxsize=0.7\hsize
\hspace{0.15\hsize}
\epsffile{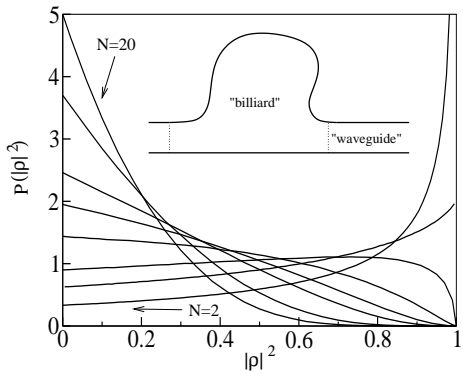}
\caption{ \label{fig:2}
Probability distribution of the phase rigidity 
$|\phase|^2$ for a wavefunction 
in an open chaotic billiard, for different numbers of propagating modes
connecting the billiard to the outside world. From bottom to top at the
left end of the figure, curves correspond to $N=2$, $3$, $4$, $6$,
$8$, $10$, $15$, and $20$. Inset:
schematic drawing of billiard and waveguides.}
\end{figure}

The joint distribution of the scattering
matrix $S$ and the Wigner-Smith time-delay matrix $Q$ of a chaotic
billiard is known from random-matrix theory \cite{kn:beenakkerRMT}:
The distribution of the proper time delays $\tau_i$ is \cite{kn:bfb1997}
\begin{eqnarray}
  P(\tau_1,\ldots,\tau_N) &=&
  \prod_{j=1}^{N} \theta(\tau_j)
  \tau_{j}^{-3N/2-1} e^{-N \tau_{\rm av}/2 \tau_j}
  \nonumber \\ && \mbox{} \times
  \prod_{i < j} |\tau_{i} - \tau_{j}|,
  \label{eq:wsdelay}
\end{eqnarray}
where $\tau_{\rm av}$ is the average delay time and
$\theta(x) = 1$ for $x > 0$ and $0$ otherwise, whereas 
the unitary matrix $U$ is uniformly distributed in the group
of unitary $N \times N$ matrices. A numerical evaluation of 
the probability distribution of the phase rigidity $|\rho|^2$
is shown in Fig.\ \ref{fig:2} for several values of $N$.
For the case $N=2$ of two single-mode waveguides, the probability
distribution $P(\phase)$ can be found in closed form,
\begin{eqnarray}
  P(\phase) &=&
  \frac{6 + 2 (1-|\phase|^2)^{-1/2}}{3 \pi (1 +
  (1-|\phase|^2)^{1/2})^3},
  \ \ 0 \le |\phase| < 1.
  \label{eq:Pp}
\end{eqnarray}
For a billiard coupled to the outside world via $N \gg 1$ channels
$P(\phase)$ becomes Gaussian,
\begin{equation}
  P(\phase) = \frac{N}{4 \pi} e^{-N |\phase|^2/4}.
\end{equation}
This is the same functional form as the
phase-rigidity distribution for a quantum dot in a large uniform
magnetic field
\cite{kn:falkoefetov1994,kn:falkoefetov1996,kn:vanlangen1997}.

Following Refs.\ 
\onlinecite{kn:zyczkowskilenz1991,kn:pnini1996,kn:saichev2002}, the joint 
distributions of intensities and current densities away from the
boundary of the cavity for one wavefunction $\psi_{\mu}$ 
can then be calculated from Berry's ansatz that $\psi_{\mu}$
can be written as a random superposition of plane waves
\cite{kn:berry1977},
\begin{equation}
  \psi_{\mu}({\bf r}) = \sum_{\vk} a_{\mu}(\vk) e^{i \vk \cdot {\bf r}}.
  \label{eq:Berry}
\end{equation}
In Eq.\ (\ref{eq:Berry}), all wavevectors $\vk$ have the same
modulus, while the amplitudes $a_{\mu}(\vk)$ are random complex
numbers. For a closed cavity, amplitudes of time-reversed plane waves
are related, $a_{\mu}(\vk) = e^{2 i \phi} a_{\mu}(-\vk)^*$,
where $\phi$ does not depend on $\vk$. For an open cavity, no such
strict relation exists, although some degree of correlation between
$a_{\mu}(\vk)$ and $a_{\mu}(-\vk)$ persists in order to ensure 
the correct value of the scalar product of $\psi_{\mu}$ and
$\psi^*_{\mu}$, {\em cf.}\ Eq.\ (\ref{eq:phase}) \cite{kn:saichev2002},
\begin{equation}
  \phase_{\mu\mu} =  \frac{\sum_{\vk} a_{\mu}(\vk) a_{\nu}(-\vk)}
  {\sum_{\vk} |a_{\mu}(\vk)|^2}. \label{eq:phasea}
\end{equation}
Taking the amplitudes corresponding to wavevectors pointing in
different directions from identical and independent distributions,
we see that Eq.\ (\ref{eq:phasea}) implies a relation between
the second moments of the amplitude distribution,
\begin{eqnarray}
  \langle a_{\mu}(\vk) a_{\mu}(-\vk) \rangle 
  &=& \phase_{\mu\mu} \langle |a_{\mu}(\vk)|^2 \rangle.
  \label{eq:var}
\end{eqnarray}
This, together with the normalization condition $\sum_{\vk} \langle
|a(\vk)|^2 \rangle = 1/A$, where $A$ is the area of the billiard, and
the central limit theorem, provides sufficient information to
determine the full distribution of the wavefunction $\psi$.

As an example, we consider the joint distribution of the normalized
intensity $\intensity({\bf r})=|\psi({\bf r})|^2 A$ and the magnitude
of 
the normalized
current density $\current = 
|{\bf j}({\bf r}')|$, ${\bf j} = (A/k) \mbox{Im}\, 
\psi^* {\bf \nabla} \psi$ at the positions ${\bf r}$ and ${\bf r}'$ 
where $k = \omega/c$.
The single-wavefunction distribution factorizes into separate
probability distributions for $\intensity$ and $\current$ that each
depend parametrically on the phase rigidity $|\phase|^2$
\cite{kn:zyczkowskilenz1991,kn:saichev2002},
\begin{eqnarray}
  P_{\phase}[\intensity({\bf r}),\current({\bf r}')] &=& \frac{8\current}
  {(1-|\phase|^2)^{3/2}}\,
  K_0\left( \frac{2\current \sqrt{2}}{\sqrt{1-|\phase|^2}} \right)
  \label{eq:ZL} 
  \\ && \mbox{} \times
  I_0\left( \frac{\intensity |\phase|}{1-|\phase|^2}\right)
  \exp\left( -\frac{\intensity}{1-|\phase|^2} \right),
  \nonumber
\end{eqnarray}
where $I_0$ and $K_0$ are Bessel functions.
When both position and frequency are varied to obtain the
ensemble average, a further average over $\phase$ is required,
\begin{equation}
  P(\intensity({\bf r}),\current({\bf r}')) =
  \int d\phase P(\phase) P_{\phase}(\intensity({\bf r}),\current({\bf r}')).
\end{equation}  
After such average, $P(\intensity,\current)$ no longer factorizes. 
The degree of correlation is measured through the
correlator
\begin{eqnarray}
  \langle \intensity({\bf r})^2 \current({\bf r}')^2 \rangle_c
  &=& - \frac{1}{2} \mbox{var}\, |\phase|^2,
\end{eqnarray}
where $\langle AB \rangle_c = \langle A B \rangle - \langle A
\rangle \langle B \rangle$ denotes the connected average.
(Since normalization implies that $\langle I({\bf r}) \rangle = 1$ for 
each wavefunction, correlators involving the first power of $I$
factorize.)
For a billiard with two single-mode waveguides, $\mbox{var}\, |\phase|^2
= 8(148 \ln 2 - 128 (\ln 2)^2 - 41)/9 \approx 0.078$, 
cf.\ Eq.\ (\ref{eq:Pp}). Similarly, we find for the correlator of
intensities
\begin{eqnarray}
  \langle \intensity({\bf r})^2 \intensity({\bf r}')^2 \rangle_c
  &=& 
  \mbox{var}\, |\phase|^2,
\end{eqnarray}
plus additional terms that describe short-range correlations.

Thus far we have studied the distribution of a single scattering
state in an open billiard. However, for a billiard that is coupled to
the outside world via, in total, $N$ propagating modes, there are
$N$ orthogonal scattering states at each frequency. In the remainder
of this letter we address the question of possible correlations
between these scattering states. 

This question can be studied using the framework of Ref.\
\onlinecite{kn:absw}, which generalizes the above considerations to
the problem of correlations between wavefunctions. Again, the starting
point is Berry's ansatz (\ref{eq:Berry}), with a different set of
amplitudes $a_{\mu}(\vk)$ for each scattering state $\psi_{\mu}$,
$\mu=1,\ldots,N$. We continue to take amplitudes $a_{\mu}(\vk)$ from
identical and independent distributions for different directions of 
$\vk$, whereas we allow for correlations
between amplitudes of time-reversed waves and between amplitudes of
different scattering states. Such correlations are necessary, because
the in-cavity parts of different scattering states and their
time-reversed states are not orthogonal, see, {\em e.g.}, Eq.\
(\ref{eq:smith}). Hence, the second moments of the amplitudes
$a_{\mu}(\vk)$ should be chosen such that
\begin{eqnarray}
  n_{\mu\nu} &\equiv& \frac{\sum_{\vk} a_{\mu}(\vk) a_{\nu}(\vk)^*}
  {(\sum_{\vk} |a_{\mu}(\vk)|^2)^{1/2} (\sum_{\vk} |a_{\nu}(\vk)|^2)^{1/2}},
  \nonumber \\ &=& 
  \frac{\int d{\bf r}\, \psi_{\mu}({\bf r})^* \psi_{\nu}({\bf r})}
  {(\int d{\bf r}\, |\psi_{\mu}({\bf r})|^2
    \int d{\bf r}'\, |\psi_{\nu}({\bf r}')|^2)^{1/2}}, 
  \label{eq:nn}\\
  \rho_{\mu\nu} &\equiv& \frac{\sum_{\vk} a_{\mu}(\vk) a_{\nu}(-\vk)}
  {(\sum_{\vk} |a_{\mu}(\vk)|^2)^{1/2} (\sum_{\vk} |a_{\nu}(\vk)|^2)^{1/2}}
  \nonumber \\ &=& 
  \frac
  {\int d{\bf r}\, \psi_{\mu}({\bf r}) \psi_{\nu}({\bf r})}
  {(\int d{\bf r}\, |\psi_{\mu}({\bf r})|^2
    \int d{\bf r}'\, |\psi_{\nu}({\bf r}')|^2)^{1/2}},
  \label{eq:rhon}
\end{eqnarray}
where, as before, the integrals are taken over the billiard only and
we have chosen the normalization such that $n_{\mu\mu}=1$.
Equations (\ref{eq:nn}) and (\ref{eq:rhon}) then impose the following
relations for second moments of the amplitude distributions:
\begin{eqnarray}
  \langle a_{\mu}(\vk) a_{\nu}(\vk)^* \rangle
  &=&
  n_{\mu\nu} \langle |a_{\mu}(\vk)|^2 \rangle, \\
  \langle a_{\mu}(\vk) a_{\nu}(-\vk) \rangle
  &=&
  \rho_{\mu\nu} \langle |a_{\mu}(\vk)|^2 \rangle.
\end{eqnarray}
Repeating the same arguments as those leading to Eq.\
(\ref{eq:rhomu}), we find that $n_{\mu\nu}$ and $\rho_{\mu\nu}$
can be expressed in terms of eigenvectors and eigenvalues of the 
time-delay matrix,
\begin{eqnarray}
  n_{\mu\nu} &=& \frac
  {\sum_{j} U_{j\mu}^* U_{j\nu} \tau_j}
  { (\sum_{j} |U_{j\mu}|^2 \tau_{j}
  \sum_{i} |U_{i\nu}|^2 \tau_{i} )^{1/2}}
  , \nonumber \\
  \rho_{\mu\nu}
  &=& \frac
  {\sum_{j} U_{j\mu} U_{j\nu} \tau_j}
  { (\sum_{j} |U_{j\mu}|^2 \tau_{j}
  \sum_{i} |U_{i\nu}|^2 \tau_{i} )^{1/2}}
  .
\end{eqnarray}
The full distribution of the complex numbers $n_{\mu\nu}$ and
$\phase_{\mu\nu}$ then follow from the known 
distributions of the $N \times N$ unitary matrix $U$ and the
proper time delays $\tau_j$, $j=1,\ldots,N$.
A simple expression is obtained in the limit $N \gg 1$, when 
$n_{\mu\nu}$ and $\rho_{\mu\nu}$ acquire a Gaussian distribution,
with zero mean and with variance given by
\begin{eqnarray}
  &&  
  \langle n_{\mu\nu} n_{\sigma \tau} \rangle
  = 
  \frac{1}{N}
  \delta_{\mu\tau} \delta_{\nu\sigma}  \ \ \mbox{if $\mu \neq \nu$},
  \nonumber \\
  && \langle \rho_{\mu\nu} \rho_{\tau\sigma}^* \rangle =
  \frac{2}{N}
  (\delta_{\mu\tau} \delta_{\nu\sigma}
  + \delta_{\mu\sigma} \delta_{\nu\tau}), \nonumber
  \\
  &&  \langle n_{\mu\nu} \rho_{\sigma\tau} \rangle = 
  \langle \rho_{\mu\nu} \rho_{\tau\sigma} \rangle
  = 0.
\end{eqnarray}

Short range correlations between different scattering modes arise
from the fact that $\rho_{\mu\nu}$ and $n_{\mu\nu}$ are nonzero for
$\mu \neq \nu$. These correlations exist both if statistics is taken
as a function of position only and if the ensemble also involves a
frequency average. For example, for the second moment of the intensity
and current density distributions, we find from Eq.\ (\ref{eq:Berry})
\begin{eqnarray}
  \langle \intensity_{\mu}({\bf r}) \intensity_{\nu}({\bf r}') \rangle_c
  &=& (|n_{\mu\nu}|^2 + |\rho_{\mu\nu}|^2) J_0(k |{\bf r}-{\bf r}'|), 
  \\
  \langle j_{\mu,\alpha}({\bf r}) j_{\nu,\beta}({\bf r}')
  \rangle &=&
  \frac{1}{4} \delta_{\alpha\beta}
  (|n_{\mu\nu}|^2 + |\rho_{\mu\nu}|^2) J_0(k
  |{\bf r}-{\bf r}'|),
  \nonumber
\end{eqnarray}
with $\alpha,\beta=x,y$.
For the case $N=2$ of a billiard with two single-mode waveguides one has
$\langle |\rho_{12}|^2 \rangle = (64 \ln 2 - 37)/15
\approx 0.49$ and $\langle |n_{12}|^2 \rangle =
(26 - 32 \ln 2)/15 \approx 0.25$.
Long-range correlations arise from the fluctuations of the
``scalar products'' $n_{\mu\nu}$ and $\rho_{\mu\nu}$ and exist 
only if the ensemble involves a
frequency average. The lowest moment with long-range correlations
is
\begin{eqnarray}
  \langle \intensity_{\mu}({\bf r})^2 \intensity_{\nu}({\bf r}')^2 \rangle_c
  &=&
  -2  \langle \intensity_{\mu}({\bf r})^2 \current_{\nu}({\bf r}')^2 \rangle_c
   \\ &=&
  \langle |\rho_{\mu\mu}|^2 |\rho_{\nu\nu}|^2 \rangle
  - \langle |\rho_{\mu\mu}|^2 \rangle
  \langle |\rho_{\nu\nu}|^2 \rangle.
  \nonumber
\end{eqnarray}
For $N=2$ one has
$\langle |\rho_{11}|^2 |\rho_{22}|^2 \rangle
  - \langle |\rho_{11}|^2 \rangle
  \langle |\rho_{22}|^2 \rangle =
  8(5792\, \ln 2 - 4480 (\ln 2)^2 - 1861)/315
\approx 0.032$.

In conclusion, we have calculated the wavefunction distribution for
wavefunctions in an open chaotic billiard for the case that the
ensemble average involves both an average over frequency and
position. Fluctuations and correlations of the phase rigidities lead
to long range correlations between intensities and current
densities. Our results are relevant for a fit-parameter free 
measurement of the real-to-complex wavefunction crossover.

\begin{acknowledgments}
We thank Karsten Flensberg for important discussions.
This work was supported by NSF under grant no.\ DMR
0086509, and by the Packard foundation.
\end{acknowledgments}

\end{document}